\documentclass[twocolumn,a4paper,showpacs]{revtex4}
\usepackage{graphicx}
\usepackage{amssymb}
\newcommand{\etal}{{\it et al}.}
\begin{document}

\title{Hydrodynamic modeling of granular flows in a modified Couette cell}

\author{Pierre Jop}
\affiliation{Laboratoire de Physique de la Mati\`ere Condens\'ee et Nanostructures, Univ. Lyon; Univ. Lyon 1;
CNRS UMR 5586, Domaine Scientifique de la Doua - F-69622 Villeurbanne Cedex, France.}
\date{\today}

\begin{abstract}
We present simulations of granular flows in a modified Couette cell, using a continuum model recently proposed for dense granular flows. Based on a friction coefficient, which depends on an inertial number, the model captures the positions of the wide shear bands. We show that a smooth transition in velocity-profile shape occurs when increasing the height of the granular material, leading to a differential rotation of the central part close to the surface. The numerical predictions are in qualitative agreement with previous experimental results. The model provides predictions for the increase of the shear bands width when increasing the rotation rate.
\end{abstract}


\maketitle
%

\begin{figure}
\includegraphics[width=8cm]{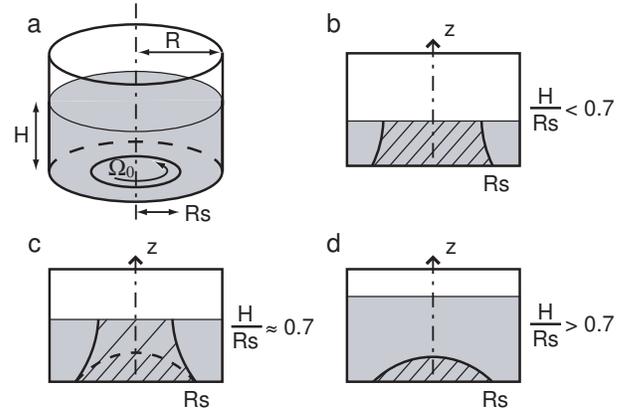}
\caption{a) Modified Couette cell: the bottom is split into a rotating disc and an outer ring at rest. From previous studies, in the stationary state, three regimes are identified depending on the filling height (the hatched areas rotate as quasi-solid bodies): (b) the open regime, (c) the intermediate regime when $H/R_s\approx0.7$ and (d) the closed system.}
\label{fig:modifcouette}
\end{figure}

Divided materials such as granular material often exhibit a localization of deformations when slowly sheared \cite{oda98,mueth00,mousse}. This behavior produces shear bands that are gene\-rally narrow, typically 10 grain-diameters wide for a granular flow. They often take place close to boundaries or at the interface between a flowing and a static part. Quasi-static motions are experimentally observed inside these bands and can lead to different velocity profiles like exponential or Gaussian, depending on the system \cite{mueth00,komatsu01}. However a theoretical framework is still lacking to understand them.
 A modified Couette configuration has recently brought the possibility of observing wide shear bands in granular media \cite{fenistein03} and allows to more deeply test recent theories \cite{unger04,depken06}. In this configuration sketched in Fig.~\ref{fig:modifcouette}(a), the bottom is split into a rotating disc (at a given angular velocity $\Omega_0$) and a ring fixed to the wall at rest. The three control parameters are the radius of the disc $R_s$, the filling thickness $H$ and the angular velocity $\Omega_0$. The experimental results are summarized here. For a small thickness compared to the disc radius, the disc drives the rotation of a column up to the free surface as shown in Fig.~\ref{fig:modifcouette}(b). However, when the height is large, only a dome rotates deeply below the surface [Fig.~\ref{fig:modifcouette}(d)]. The transition occurs around the ratio $H/R_s=0.7$ \cite{fenistein03,fenistein04,cheng06}, where two shear bands exist [Fig.~\ref{fig:modifcouette}(c)]. In this regime, a differential rotation (called precession in the following) appears: the central upper part rotates more slowly than the bottom \cite{fenistein06,cheng06}. A theoretical model, which is based on a principle of minimization of the Coulomb friction dissipation, 
 predicts infinitely sharp shear band and a hysteretic transition between open and closed regimes \cite{unger04}. T\"or\"ok \etal~recently introduce a random disorder, they  find a smooth transition and wide shear bands with corrects scaling laws when averaging over an ensemble of sharp bands \cite{torok06}.
 
In this paper, we present an alternative approach. A constitutive law has been proposed, 
 which quantitatively describes some dense surface flows and instabilities \cite{jop06,forterre06,jop07}. Could this hydrodynamic model describe the flow in the modified Couette cell? What are the shapes of bands and is there a smooth transition?

In the local model proposed in \cite{jop06}, the granular media is described as an incompressible fluid governed by a visco-plastic constitutive law. Assuming an ensemble of grains of diameter $d$ and density $\rho_s$, the shear stress $\sigma_{ij}$ is related to the strain rate $\dot\gamma_{ij}$ as follows:
\begin{equation}
\sigma_{ij}=-P\delta_{ij}+\tau_{ij} { \rm  \hspace{4 mm} with \hspace{4 mm}}\tau_{ij}=   \frac{\mu(I)P}{|\dot{\gamma}|}\dot{\gamma}_{ij}\equiv\eta\dot{\gamma}_{ij},
\label{eq:rheol}
\end{equation}
where $P$ is the isotropic pressure, $\mu(I)$ is a friction coefficient, which depends on the inertial number $I$  ($I~=~\frac{|\dot \gamma | d}{\sqrt{P/\rho_s}}$), the norm $|X|$ corresponds to $\sqrt{X_{ij}X_{ij}/2}$ and $\eta$ is the effective viscosity. The friction coefficient starts from a minimum value $\mu_s$ and increases with $I$ (see Ref. \cite{jop06} for details about the rheology). Although this model was not designed to reproduce quasi-static creeping flows (the constitutive law ensures the existence of a finite yield stress $|\tau|>\mu_sP$), we expect it could contribute to understand what kind of phenomenon underlies the properties of shear bands in this geometry.
 We assume the flow to be axisymmetric and compute the axial velocity $u_\theta(r,z)=r\Omega(r,z)$, only in a radial section of the cell, governed by the following equations: 
\begin{equation} 
\rho\frac{\partial u_\theta}{\partial t}=\frac{1}{r^2}\frac{\partial}{\partial r}\left ( \eta r^2\frac{\partial u_\theta}{\partial r}-\eta r u_\theta \right ) + \frac{\partial}{\partial z}\left ( \eta \frac{\partial u_\theta}{\partial z}\right),
\label{eq:rheolE}
\end{equation}
\begin{equation}
\rho\frac{u_\theta^2}{r}=\frac{\partial P}{\partial r} { \rm  \hspace{3 mm} and \hspace{3 mm}}
0=\rho g -\frac{\partial P}{\partial z}.
\label{eq:pressure}
\end{equation}
The simulations are made on a fixed grid (72 $\times$ 72). Using a finite-difference scheme, we integrate the Cauchy equation for $\Omega$ from the material at rest until a steady state is reached with the following boundary conditions: a stress-free condition at the free surface and a no-slip-velocity condition at the wall and at the bottom (the angular velocity is then given by $\Omega|_{(z=0)}=\Omega_0$ for $r<R_s$ and $0$ for $r>R_s$ \footnote{In previous studies, the inner disk is at rest while the outer anneal rotates. We have checked that even in the worst case the resulting profiles differ less than 2\% due to the radial acceleration.}). The set of parameters used in simulations is similar to the experimental one found in Ref. \cite{fenistein06}. To simplify computations, we assume moreover the free surface to remain flat and horizontal: for low angular velocities the elevation is negligible and would reach $5\%$ of the radius for a viscous fluid when $\Omega_0$ is multiplied by 20. The pressure is thus the sum of a hydrostatic part and an inertial part due to the rotation.

\begin{figure}
\includegraphics[width=8.6cm]{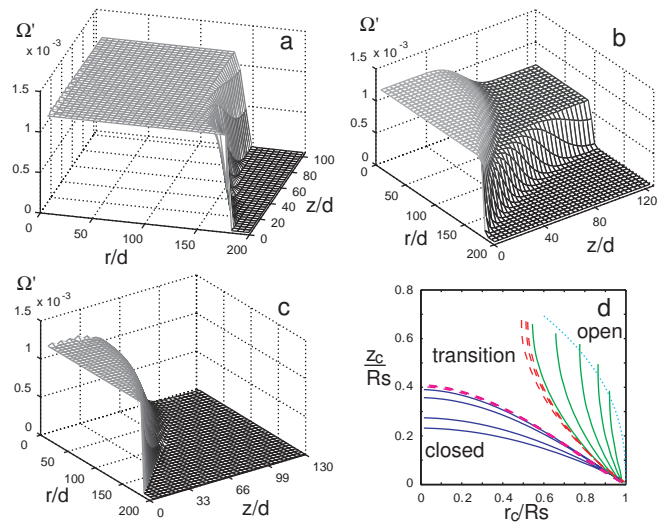}
\caption{a-c) Stationary angular velocity profiles in a radial section of the Couette cell for $R_s=180d$ ($\Omega'=\Omega/(g/d)^{1/2}$). a) $H$~=~$105d$ ($H/R_s$~=~$0.59$), the rotation reaches the free surface (open regime). b) $H=126.5d$ ($H/R_s=0.706$), this case corresponds to the transitory regime. c) $H=130d$ ($H/R_s=0.73$), above the transition we observe the closed regime. d) Position of the shear bands for different aspect ratio. Solid lines: $H/R_s=$ 0.39, 0.5, 0.59, 0.63, 0.66 (open systems), 0.73, 0.78, 0.84 (closed systems). Dashed lines: $H/R_s=$0.700, 0.706, 0.714 (intermediate regime). Dotted line: experimental position of the shear band at the free surface from Ref.~\cite{fenistein04}.}
\label{fig:profils3D}
\end{figure}

Figures \ref{fig:profils3D}(a-c) show typical angular-velocity profiles in the radial section of the cylinder for three different filling heights. For $H/d= 105$ ($H/R_s = 0.59$), the central part, which rotates, emerges at the surface of the granular material [Fig.~\ref{fig:profils3D}(a)]. For $H/d = 130$ ($H/R_s = 0.73$) in Fig.~\ref{fig:profils3D}(c), the rotating zone is reduced to a cupola near the rotating disc. Close to the transition [Fig.~\ref{fig:profils3D}(b)], we observe two shear bands ($H/R_s=0.706$): three quasi-rigid bodies are separated by narrow shear bands. The angular velocity of the upper central part is in-between 0 and $\Omega_0$. A first conclusion of this study is that the continuum model qualitatively reproduces the experimentally observed phenomenon. Can it also describe the shape and the width of these shear bands? %
To study the shape, we define the position of shear bands in the bulk by the points where the angular velocity equals the average value of neighboring zones. Figure~\ref{fig:profils3D}(d) shows the positions of the shear bands $z_c(r_c)$ for different filling heights. The continuous lines correspond to the extreme regimes. For increasing heights, the open shear band is more curved and is closer to the center. When the transition occurs, its radius at the free surface seems to reach a limit and the open shear band vanishes while the closed one appears below the surface. Both coexisting shear bands are shown by the dashed lines in Fig.~\ref{fig:profils3D}(d).
The dotted line corresponds to the empirical scaling of the shear position at the free surface (in the open state) from experiments of Fenistein \etal~\cite{fenistein04}. It is in good agreement with the curve tips of our simulations, but a slight departure can be observed as obtained by the Unger \etal's model \cite{unger04}.

\begin{figure}
\includegraphics[width=5.7cm]{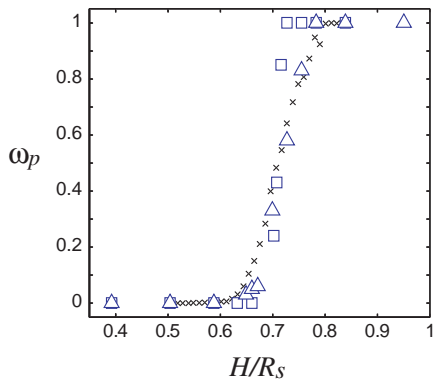}
\caption{Precession of the surface core: normalized difference of angular velocity $\omega_p=1-\Omega/\Omega_0|_{z=H,r=0}$ as a function of the rescaled height $H/R_s$. Comparison between simulations for two angular velocities ($\Omega_0=0.16$~rad.s$^{-1}$~$\Box$ and $\Omega_0=3.2$~rad.s$^{-1}$~$\bigtriangleup$) and experimental results from Ref.~\cite{fenistein06}~($\times$).}
\label{fig:precession}
\end{figure}

Let us consider in more details the intermediate regime, where our simulations reveal the existence of two shear zones bordering three rigid parts: the bottom rotates at $\Omega_0$, the lateral part is at rest and, in between, the upper central part rotates at a smaller angular velocity. To analyze the evolution of this phenomenon and to compare it with the experimental one \cite{fenistein06}, we plot in Fig.~\ref{fig:precession} the normalized difference between the angular velocity at the bottom and at the free surface on the symmetric axis: $\omega_p=1-\Omega/\Omega_0|_{z=H,r=0}$. For a given angular velocity, we obtain first that the transition between open and closed systems is continuous as observed in experiments \cite{fenistein06} and DEM simulations \cite{cheng06,lechman05}. We do not find any hysteretic phenomenon unlike the predictions of Unger's model \etal~\cite{unger04}. 
However, the transition, which occurs around $H/R_s\approx0.7$, is sharper in our simulations than in experiments, although the simulated curve (squares) and the Fenistein \etal's experiments (crosses) correspond to the same set of parameters. This could be due to the inability of our model to capture the quasi-static properties due to non-local effects \cite{gdrmidi04}. 
The important point is thus that we can reproduce the smooth transition using a shear-rate-dependent friction coefficient. As a consequence, the observed precession is the result of the superposition of two shear bands. When the rotation rate is increased by a factor 20 ($\Omega_0=3.2$~rad.s$^{-1}$), the transition zone broadens as shown by the precession ratio $\omega_p$ in Fig.~\ref{fig:precession} (triangles). We can notice moreover that, if the radial acceleration is negligible (left part of the equation \ref{eq:pressure}), equations governing the model are invariant when all lengths are normalized by $R_s$ and velocities are divided by $R_s^{3/2}$. The latter exponent comes from the expression of $I$. The rescaled variables are then $H/R_s$, $W/R_s$ and $\Omega/R_s^{1/2}$. 
The dependence on $\Omega_0$ can be translated in terms of $R_s$:  for very large systems, the extent of transition should decrease for increasing $R_s$. This prediction is in agreement with T\"or\"ok \etal's model \cite{torok06}, which predicts an infinitely sharp transition for $R_s\rightarrow\infty$.

\begin{figure}
\includegraphics[width=8.6cm]{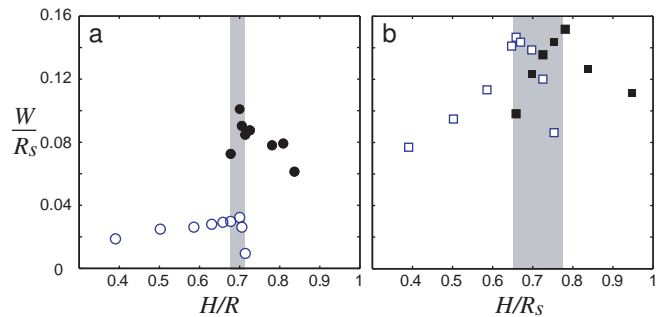}
\caption{Evolution of shear band widths as a function of the aspect ratio through the transition for two different angular velocities: a) $\Omega_0=0.16$~rad.s$^{-1}$~$\circ$. b) $\Omega_0=3.2$~rad.s$^{-1}$~$\Box$. White (resp. black) symbols correspond to open (resp. closed) shear bands and the gray regions to the transition zone ($0.03<\omega_p<0.97$).}
\label{fig:widthhoverrs}
\end{figure}

We now focus on the width of shear bands at the free surface and in the bulk, first for a low angular velocity ($\Omega_0$=0.16~rad.s$^{-1}$). In the following, we will extract the band widths from the angular-velocity profiles in the surface and along the symmetric axis in the bulk fitted by an error function: $\Omega/\Omega_0=1/2+1/2\rm{erf}[(X_c-x)/W]$ (The axial shear band is very well fitted whereas the surface velocity profile is more linear). 
Following the dimensional analysis, we use the scaling $W/R_s$. The width of the open shear band, whose evolution with the filling height is shown in Fig.~\ref{fig:widthhoverrs}(a), is very narrow. The width $W$ increases with $H/R_s$ (white circles) and suddenly decreases below the resolution of the grid close to the transition (gray region).  
After the transition, we plot the width of the vertical shear band along the axis of the cylinder. The vertical width in the closed system (black circles) is here larger and decreases with the height. Comparing with experimental and numerical results  \cite{fenistein06,cheng06,torok06}, we can notice that our open shear bands are very narrow but that the width of closed shear bands is of the same magnitude\\ 
For a higher angular velocity ($\Omega_0$=3.2 rad.s$^{-1}$), in addition to the enlargement of the transition range, both radial and vertical shear bands widen [Fig.~\ref{fig:widthhoverrs}(b)].  The global behavior is similar to the one at low velocity, except that in that case, the variations of the widths within the transition state are more obvious: the widths of each band go to zero as they disappear. These variations of widths can be understood as follows: the rheology reduces to a Coulomb friction law for small $I$ (the normalized shear rate), so when the velocity difference between two neighbor zones decreases, the friction coefficient decreases and the shear bands become narrower. 
These modifications of widths with respect to $\Omega_0$ are in contrast with what is observed in direct simulations and in experiments where the velocity pattern is constant over a given range of rotation rate \cite{fenistein04,cheng06}. Before concluding on this point, we study in more details the influence of the driven velocity on the width of the shear bands.

\begin{figure}
\includegraphics[width=6.5cm]{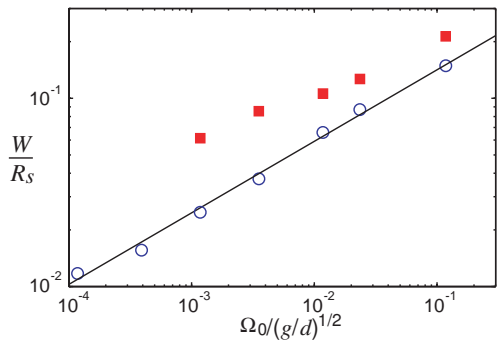}
\caption{Evolution of the width of a closed ($H/R_s=0.84$ {\tiny$\blacksquare$}) and open ($H/R_s=0.50$ $\circ$) shear band as a function of angular velocity $\Omega_0$. The data are fitted by a power law of exponent 0.38 (line). For the highest velocity and the open regime, the assumption of a flat free surface is no more valid.}
\label{fig:widthomega}
\end{figure}

In the following paragraph, we focus on two given aspect ratios: an open shear band ($H/R_s=0.5$) and a closed shear band ($H/R_s=0.84$). First, we observe that the positions of shear bands are quite unchanged when increasing the rotation: the position of the open one moves very slightly toward the center. Second, we study the evolutions of widths with respect to the angular velocity as shown in Fig.~\ref{fig:widthomega}. For the open shear band, the width increases with a power law of angular velocity which exponent is $\alpha=0.38$. This relation means first that, in the limit of very low angular velocity, this shear band is infinitely narrow. This is consistent with the fact that our model reduces to  a simple Coulomb friction law (a Drucker-Prager criterion) for I$\rightarrow$0. In Unger \etal's model \cite{unger04}, a constant friction law produces sharp shear bands. 
Concerning the closed shape, the widths have roughly the same behavior, but for low rotation rate we observe a slight departure from the power law. This behavior could be attributed to numeric artifact that vanishes when the size of the mesh grid decreases. Since this numerical effect does not alter the open shear band, we assume that the geometry could influence in different ways these two shear bands.
 From the dimensional analysis, the scaling for the radial width reported in previous papers $[W/R_s^{2/3}=f(H/R_s)]$ \cite{fenistein06,torok06} can not be derived: First we find that $W/R_s$ is the relevant variable for the closed regime in the inertial regime, second the observed scaling law for the open shear band, $W/R_s\propto\Omega_0^\alpha$ (i.e. $\propto\Omega_0^\alpha/R_s^{\alpha/2}$), implies instead that $W/R_s^{\beta}$ is a function of $H/R_s$ for a given $\Omega_0$, with $\beta=1-\alpha/2\approx0.84$. This discrepancy with previous results could come from the inability of the model to reproduce quasi-static phenomena, which govern the precise shape of shear bands. However, this dependance on the rotation rate might be seen experimentally for higher angular velocities ($\left<I\right>>3~10^{-3}$ for $\Omega_0>0.02$~$\sqrt{g/d}$).

 
In conclusion, we have shown that a continuum model based on a hydrodynamic approach allows the computation of the main properties of shear bands in the modified Couette geometry: the threshold, which corresponds to a 3D generalization of the Coulomb friction threshold ($|\tau|<\mu_sP$), is sufficient to determine the position of shear bands, and the experimentally observed transition in shapes is qualitatively reproduced. Second, we predict a smooth transition, which differs from the first-order transition predicted by Unger \etal's model. This comes from the increase of $\mu$ with the inertial number inside the shear band and from the superposition of two shear zones. 
 Like in the plane-Couette geometry without gravity, the model predicts wide shear bands as soon as $\mu$ depends on $I$. However, the correct widths are not predicted, specially the width of the radial shear band. This problem arises mainly from the quasi-static properties of the flow, but when we leave the quasi-static regime and enter in the dense regime (locally $I\approx 0.05$), 
 some predictions, which can be tested experimentally, are made concerning the evolution of these bands with respect to $\Omega_0$. In experiments at low shear, the effects of $I$ may be hidden by other phenomenon. Another parameter, such as the one used in the Depken \etal's shear-free-sheet model \cite{depken06}, may selects the properties of the velocity profiles or the assumption of an isotropic pressure, which seems valid in rapid flows, may fail as it does in static piles.





The author thanks O. Pouliquen, Y. Forterre for discussions and carefully reading the manuscript and also M. van Hecke for useful discussions.

\end{document}